\documentclass[9pt,twocolumn,twoside]{osajnl}

\journal{ol} 

\setboolean{shortarticle}{true}

\title{Calibration-free quantitative phase imaging in multi-core fiber endoscopes using end-to-end deep learning}

\author[1]{Jiawei Sun}
\author[1,2,*]{Bin Zhao}
\author[1]{Dong Wang}
\author[1]{Zhigang Wang}
\author[1,3]{Jie Zhang}
\author[3,4]{Nektarios Koukourakis}
\author[3,4]{Jürgen W. Czarske}
\author[1,2,$\dag$]{Xuelong Li}

\affil[1]{Shanghai Artificial Intelligence Laboratory, Longwen Road 129, Xuhui District, 200232 Shanghai, China}
\affil[2]{School of Artificial Intelligence, OPtics and ElectroNics (iOPEN), Northwestern Polytechnical University, Xi’an, China}
\affil[3]{Laboratory of Measurement and Sensor System Technique (MST), TU Dresden, Helmholtzstrasse 18, 01069 Dresden, Germany}
\affil[4]{Competence Center for Biomedical Computational Laser Systems (BIOLAS), TU Dresden, Dresden, Germany}

\affil[*]{binzhao111@gmail.com}
\affil[$\dag$]{li@nwpu.edu.cn}

\begin{abstract}
Quantitative phase imaging (QPI) through multi-core fibers (MCFs) has been an emerging in vivo label-free endoscopic imaging modality with minimal invasiveness. However, the computational demands of conventional iterative phase retrieval algorithms have limited their real-time imaging potential. We demonstrate a learning-based MCF phase imaging method, that significantly reduced the phase reconstruction time to 5.5 ms, enabling video-rate imaging at 181 fps. Moreover, we introduce an innovative optical system that automatically generated the first open-source dataset tailored for MCF phase imaging, comprising 50,176 paired speckle and phase images. Our trained deep neural network (DNN) demonstrates robust phase reconstruction performance in experiments with a mean fidelity of up to 99.8\%. Such an efficient fiber phase imaging approach can broaden the applications of QPI in hard-to-reach areas.

\end{abstract}

\setboolean{displaycopyright}{true}

\begin{document}

\maketitle

Fiber endoscopes have emerged as a vital tool for high-resolution microscopic imaging in hard-to-reach areas. In contrast to conventional endoscopes with a typical diameter of several millimeters, fiber endoscopes, which could be sub-millimeter thin and flexible \cite{porat2016widefield,borhani2018learning,kuschmierz2021ultra,sun2022quantitative,badt2022real}, can pass through the organ's intricate pathways without causing harm inside the body \cite{wen2023single}, making them particularly suitable for procedures requiring utmost precision and minimal invasiveness. The reduced size and adaptability of fiber endoscopes ensure less discomfort for the patient, leading to quicker recovery times and a lower risk of complications \cite{li2021memory}. 

\begin{figure*}[t]
    \centering
    \includegraphics[width=11cm]{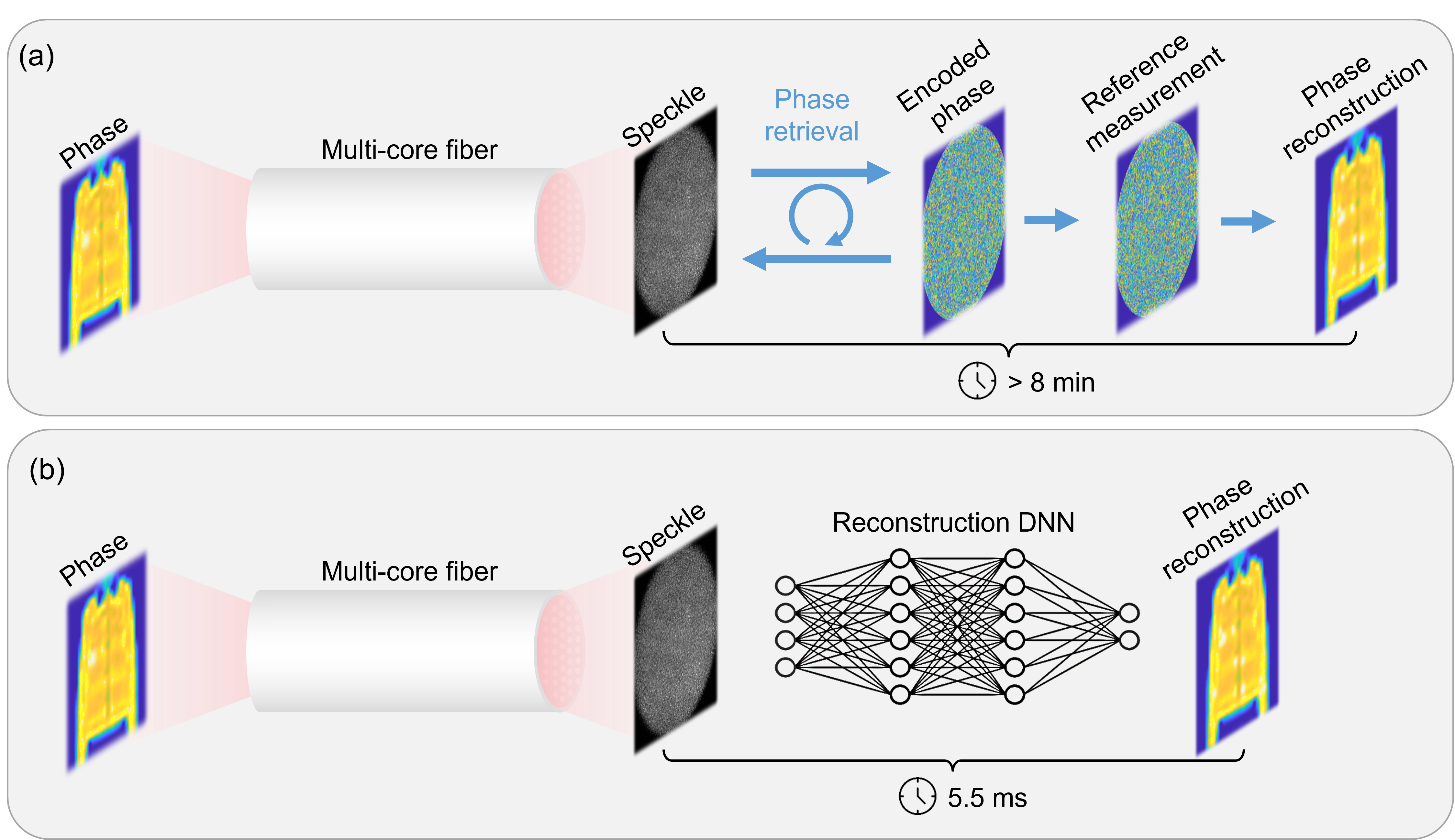} 
    \caption{(a) Conventional multi-core fiber phase imaging using iterative phase retrieval algorithms. The complete computation process including calibration takes several minutes. (b) Calibration-free multi-core fiber phase imaging based on end-to-end deep neural network. The phase reconstruction time is significantly shortened to a few milliseconds on the same computing platform.}
    \label{fig:principle}
\end{figure*}

Typical fiber endoscopes employ fluorescent imaging technique to enhance the image contrast \cite{wu2022learned}, however, the staining process could potentially introduce harmful agents to tissues. On the other hand, label-free intensity imaging avoids the risk of tissue toxicity but often falls short in terms of image contrast. QPI has proven to be a powerful tool for 3D surface imaging and enhances the contrast of cells and tissues without staining. Also, biophysical properties of biological samples like refractive index and dry mass \cite{park2018quantitative,sun2023compressive} can be derived using QPI. Nevertheless, it was challenging to achieve QPI through fiber endoscopes, primarily because of the phase distortion in multi-core fibers (MCF) \cite{kuschmierz2018self}. As demonstrated in Fig.~\ref{fig:principle}a, we previously introduced a non-interferometric speckle reconstruction method termed Far-field Amplitude-only Speckle Transfer (FAST)" to achieve QPI through an MCF \cite{sun2022quantitative}. We found that even if the phase information becomes distorted after the light field travels through the MCF, we can still retrieve both amplitude and phase details of the light field from the speckle image obtained at the measurement side using FAST. The FAST algorithm provides high-fidelity phase reconstruction from the speckle images, however, the computation effort of the FAST algorithm is relatively high due to its iterative process. Specifically, it takes more than 8 minutes to reconstruct a single phase image, including the calibration process. In clinical application scenarios, it is imperative that endoscopic imaging techniques facilitate real-time, calibration-independent imaging. This highlights the need for an alternative methodology capable of directly extracting the quantitative phase information from the speckle image captured at the measurement side of the MCF.

Recent advancements have adopted deep learning techniques to expedite the QPI image reconstruction process \cite{wu2021dnn,jo2018quantitative}. Moreover, extant literature indicates the potential of decrypting an encoded phase directly from speckle images utilizing deep learning, although only in simulated environments \cite{hai2019cryptanalysis}. This demonstrates the theoretical possibility of reconstructing the original phase directly from speckle images using deep learning for MCF phase imaging, however, networks trained on simulated data can hardly achieve accurate phase reconstructions in real-world optical systems. Furthermore, a dedicated experimental dataset tailored for fiber endoscopic phase imaging is currently not available, which is essential for effective network training. It is also challenging to record a large number of paired speckles and phase images simultaneously in one optical system.

To tackle this problem, we present an innovative optical system using holographic display technology to automatically generate 50,176 paired images in experiments, serving as a robust foundation for training the phase reconstruction DNN. Owing to the experimentally generated training dataset, the trained DNN achieves high fidelity QPI in experiments with a high speed of 5.5 ms, shown in Fig.~\ref{fig:principle}b. Such a significant enhancement in phase reconstruction speed can further improve the temporal resolution of fiber-based phase imaging. This advancement facilitates real-time calibration-free phase imaging with the fiber endoscope, strengthening its potential for real-world clinical applications.

\begin{figure}[h!]
    \centering
    \includegraphics[width=8.5cm]{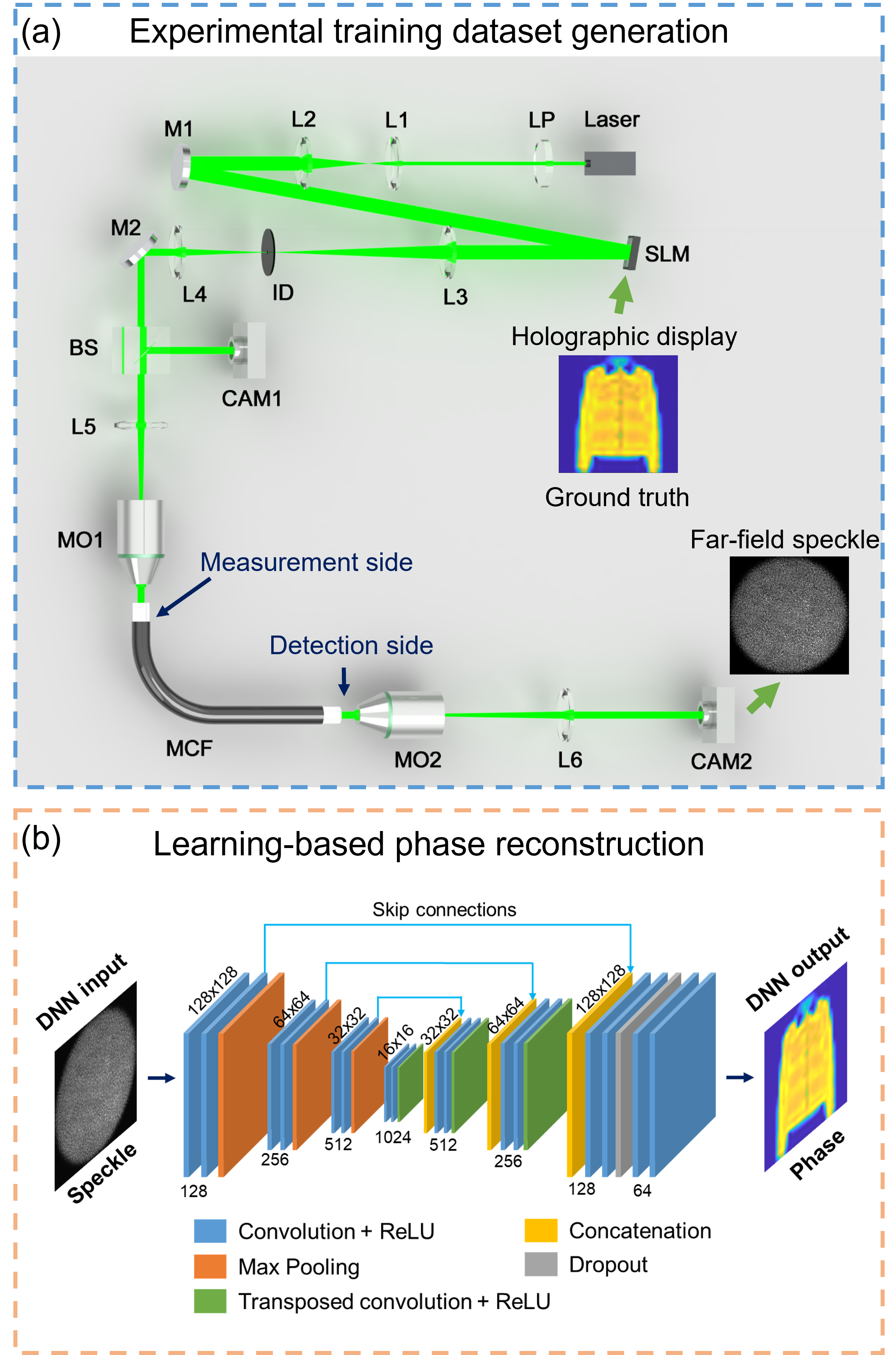} 
    \caption{(a) Scheme of the optical system for generating the training images. The ground truth phase images are displayed on the SLM and projected on the MCF facet at the measurement side. The SLM and the detection camera (CAM2) are synchronously triggered to record the corresponding paired speckle images. LP, linear polarizer; L1-L6, achromatic lenses; M1-2, mirrors; ID, iris diaphragm; BS, beamsplitter; CAM1, alignment camera; MO1-2, microscope objectives. (b) The architecture of the deep neural network for reconstructing the phase information from the speckle images.}
    \label{fig:setup}
\end{figure}

The optical system used for generating the training dataset is demonstrated in Fig.~\ref{fig:setup}a. We take images from both the MNIST fashion dataset \cite{xiao2017fashion} and the handwritten digits database \cite{lecun1998gradient}, and transform them into single precision phase images within the [0 $\pi$] range. These phase images are then adjusted to a resolution of $980\times980$ pixels and zero-padded to $1920\times1080$ pixels to fit the display resolution of the phase-only spatial light modulator (SLM) (PLUTO, Holoeye Photonics). The corresponding computer-generated holograms (CGHs) are then calculated and displayed on the SLM for precise holographic display of the ground truth phase image. The modulated laser beam (Verdi, Coherent Inc.), which is expanded by a telescope system (L1-2) for homogeneous illumination on the SLM, passes through a spatial filtering system (L3-L4, ID) to get rid of the unwanted diffraction orders. The phase image is then projected on the proximal facet of the MCF (FIGH-350S; Fujikura) at the measurement side through the microscope system (L5, MO1; $10\times$ plan achromat objective, 0.25 NA, Olympus). The incident beam is partially reflected by the MCF facet at the measurement side and projected onto the alignment camera (CAM1; uEye LE, IDS), facilitating precise alignment of the holographic display plane and the fiber facet. The far-field speckle is imaged on the detection camera (CAM2; uEye CP, IDS) through another microscope system (L6, MO2; $10\times$ plan fluo objective, 0.3 NA, Nikon). The SLM and the detection camera are synchronously triggered, which means when an image is projected onto the SLM, the camera is concurrently activated to capture the corresponding speckle image. This ensures accurate correspondence between the displayed phase images and captured speckles. The distance between the far-field image plane and the MCF facet at the detection side is 0.5 mm. 29,379 preprocessed phase images from MNIST handwritten digits database and 20,796 preprocessed phase images from the fashion MNIST database are holographically displayed on the MCF and the paired speckles are recorded as the training dataset. The input images of the network are the far-field speckle patterns captured on the detection camera. The speckle patterns and their corresponding label phase image are paired and resized to 128x128 pixels in the training dataset. For the handwritten digits, we split it into 26,001 training images, 3,300 validation images, and 78 test images. For the fashion MNIST, we use 18,001 paired images for training, 2,700 for validation, and 96 for testing.

\begin{figure}[t]
    \centering
    \includegraphics[width=8.5cm]{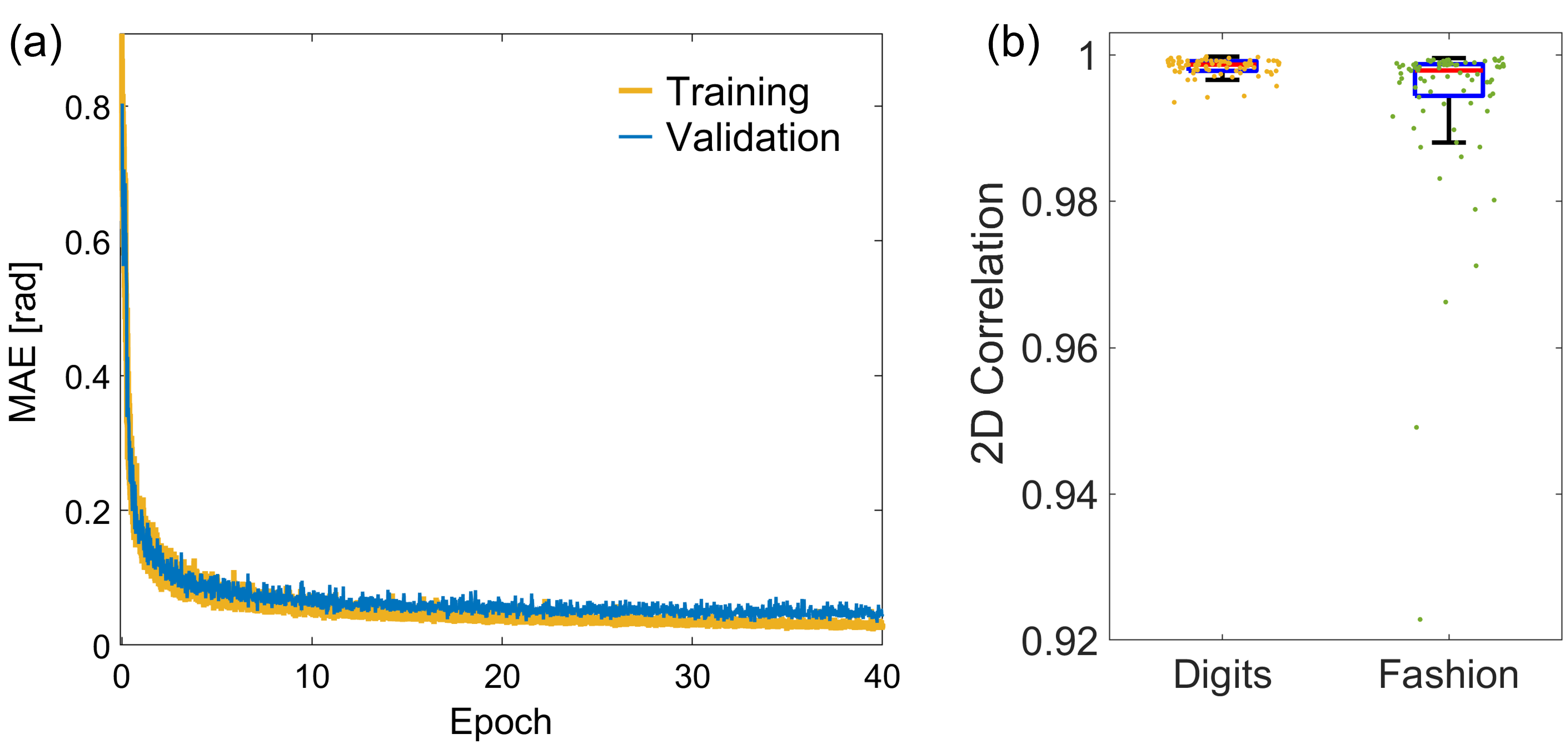} 
    \caption{(a) Training loss and validation loss during the training process. (b) Boxplot of the 2D image correlation coefficients between the DNN predicted phase images and the ground truth in the test data group of handwritten digits and fashion clothes.}
    \label{fig:train}
\end{figure}

To achieve real-time phase reconstruction through the MCF, we implement a DNN designed to derive the phase image directly from the far-field speckles. The architecture of the network is demonstrated in Fig.~\ref{fig:setup}b, which is a combination of convolutional U-Net \cite{ronneberger2015u} and ResNet \cite{he2016deep} architectures. This structure incorporates three downsampling blocks paired with three upsampling blocks with skip connections. The skip connections streamline the forward pass, allowing for direct paths through the network, which can enhance the speed of predictions. Meanwhile, the U-Net's structure reduces the number of necessary operations since it directly combines features from different network depths without the need for redundant calculations. In each sampling block, there are two convolutional layers and rectified linear unit (ReLU) layers, and the depth of the layers is marked below the block in Fig.~\ref{fig:setup}b. Furthermore, dropout layers are introduced just before the final convolutional output layers, and serves as a regularization technique within this DNN. During the training process, these layers randomly omit or deactivate certain neurons, meaning they temporarily remove them from the network. By this way, the network becomes less reliant on any individual neuron. This approach pushes the network to distribute its learned features more evenly across all neurons to avoid overfitting, leading to more generalized and robust representations \cite{srivastava2014dropout}. To quantify the difference between the predicted output and the actual output, we employed the mean absolute error (MAE) as our loss function, mathematically defined as ${MAE = \frac{1}{M \cdot N^{2}}
     \sum ^{M}_{k=1} \sum ^{N}_{i=1} \sum ^{N}_{j=1} \left| (Y_{i,j}-X_{i,j}) \right|,}$ where $Y_{i,j}$ is the label phase image, $X_{i,j}$ is the network output image, and ${i}$ and ${j}$ are the indices of the images. ${N \times N}$ indicates the image size in pixels and ${M}$ is the training mini-batch size. The weights and biases in the network are updated using adaptive moment estimation (Adam) optimizer \cite{kingma2014adam} at the end of each training mini-batch. Training was conducted on a robust computing platform equipped with an Nvidia RTX A6000 GPU and an AMD Ryzen 9 3950X CPU. Using our experimentally generated training and validation dataset, which comprises 50,002 paired images, each epoch of training took approximately 4.5 minutes. As illustrated in Fig.~\ref{fig:train}a, the MAE of the validation data reached a convergence point after 40 epochs. The average MAE was observed to be 0.045 for the fashion dataset and 0.015 for the digits dataset. To fine-tune our training regimen, we initiated with a learning rate of 0.0001, subsequently decreasing it by half every 20,000 iterations. This adaptive learning rate strategy ensures a more optimized and efficient learning process. Additionally, to enhance training efficiency, we utilized a minibatch size of 64 during the training iterations.

The phase reconstruction fidelity of the trained DNN is evaluated quantitatively using the 2D correlation coefficient between the DNN reconstructed image and the ground truth. As shown by the red line in Fig.~\ref{fig:train}b, the mean fidelity of the DNN-based phase reconstruction is 0.994 for 96 handwritten digits and 0.998 for the 78 images from fashion MINIST dataset, indicating high fidelity phase reconstruction performance in general. The fashion MNIST dataset, with its diverse and complex images, shows a greater variation in reconstruction fidelity. Specifically, the most challenging fashion MNIST test image can still achieve a decent reconstruction fidelity of 0.923, as seen in Fig.~\ref{fig:train}b.

\begin{figure}[ht]
    \centering
    \includegraphics[width=8.5cm]{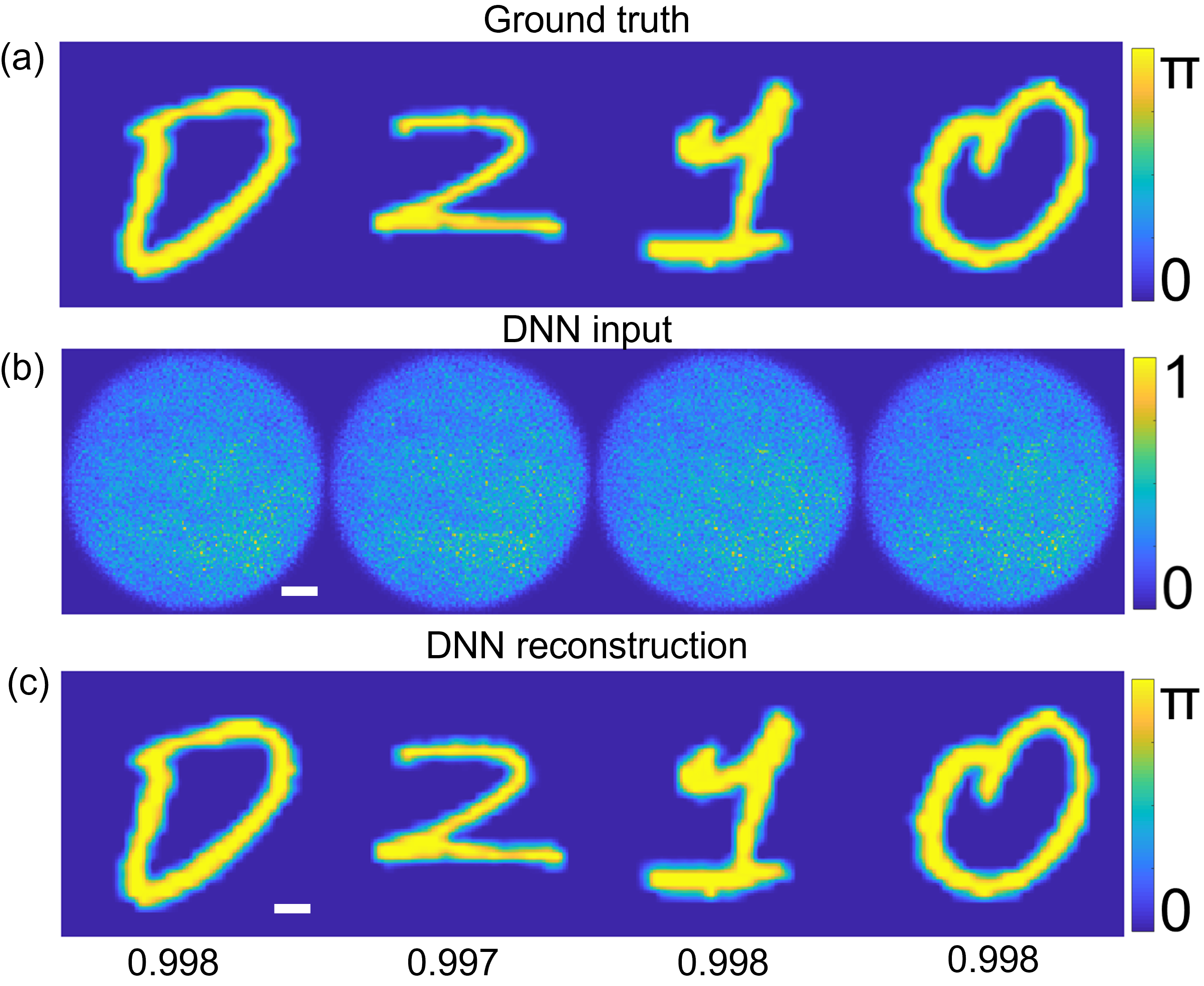} 
    \caption[Validation of the phase reconstruction DNN using MNIST handwritten digits dataset]{Phase reconstruction of handwritten digits through the MCF using the trained DNN. (a)  Ground truth. (b) Speckle images that are used as the DNN input. (c) DNN reconstructed phase images. The numbers below indicate the reconstruction fidelity characterized by the correlation coefficient. Scale bars 50~µm.}
    \label{fig:digits}
\end{figure}

To visualize the phase reconstruction performance of the DNN in experiments, test phase images of handwritten digits illustrated in Fig.\ref{fig:digits}a, are holographically projected to the MCF facet at the measurement end. When the light field travels through the MCF, its phase is distorted due to the intrinsic optical path differences between fiber cores. The corresponding far-field speckles captured by the detection camera are demonstrated in Fig.\ref{fig:digits}b. These detected speckle images serve as the network's input to test the trained DNN. As shown in Fig.~\ref{fig:digits}c, the DNN accurately reconstructs the original incident phase images from the acquired speckle images.

\begin{figure}[!ht]
    \centering
    \includegraphics[width=8.5cm]{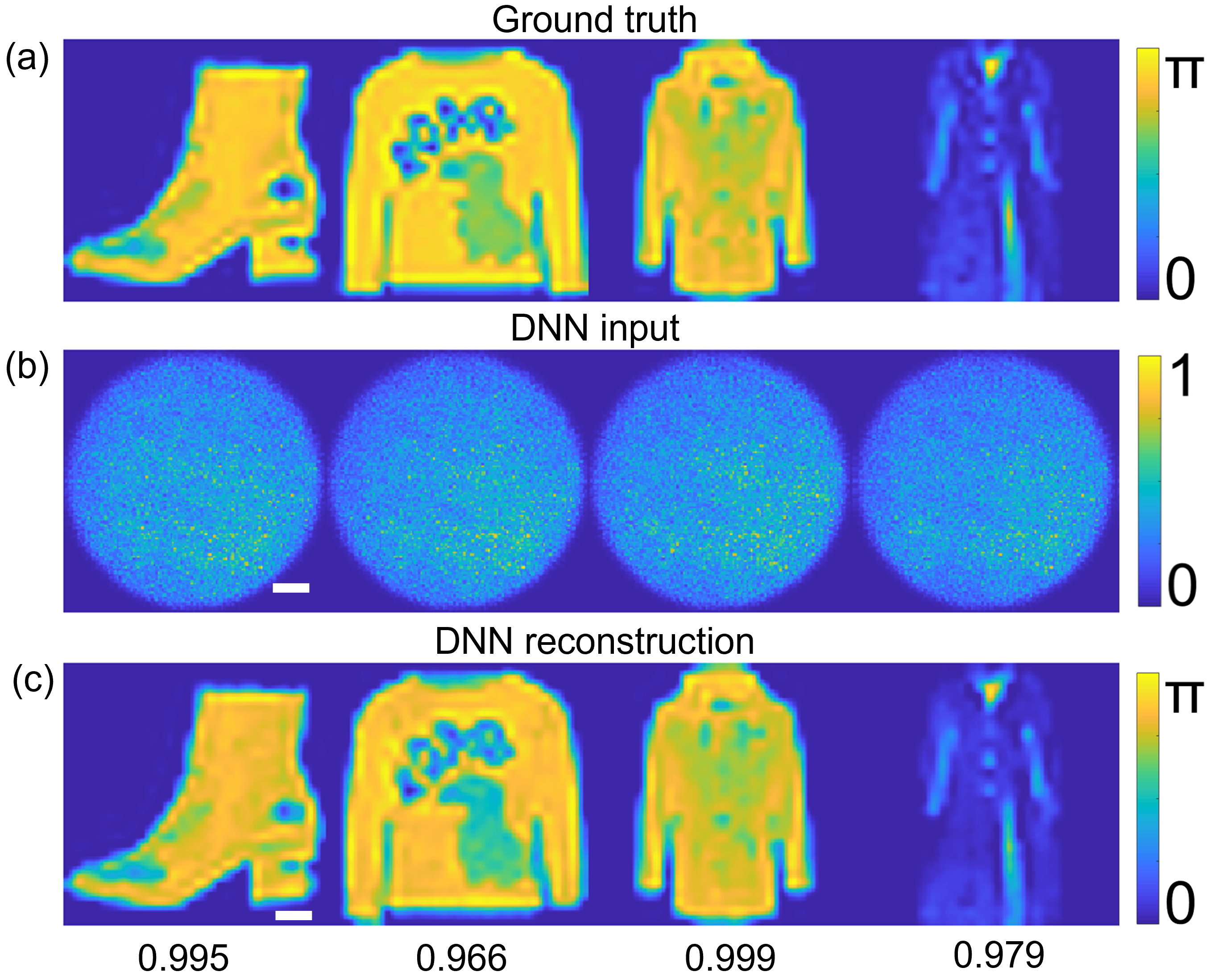} 
    \caption[Performance of the PhaseUnet for fashion MNIST dataset]{Phase reconstruction of MNIST fashion images through the MCF using the trained DNN. (a)  Ground truth phase image displayed on the SLM. (b) Speckle images that are used as the DNN input. (c) DNN reconstructed phase images. The numbers below indicate the reconstruction fidelity characterized by the correlation coefficient. Scale bars 50~µm.}
    \label{fig:fashion}
\end{figure}

It has been validated that the trained network can achieve phase reconstruction from speckle images for handwritten digits with excellent fidelity. Nevertheless, the MNIST handwritten digits are relatively simple structures, to challenge the network further and ensure its robustness in more intricate scenarios, the fashion MNIST dataset is employed. In this experiment, phase images from the fashion MNIST test dataset are projected onto the MCF's measurement end as the ground truth, see Fig.\ref{fig:fashion}a. The experimentally measured speckle images at the detection side of the MCF are demonstrated in Fig.~\ref{fig:fashion}b. To appraise the DNN's reconstruction capabilities, these detected speckle patterns are input into the trained network. Remarkably, as shown in Fig.~\ref{fig:fashion}c, the DNN successfully reconstructs the phase images of various items from their corresponding speckle images. Such proficient reconstruction underscores the versatility and robustness of the DNN in dealing with complex patterns. This capability is crucial when considering real-world applications where data can be intricate and varied. The successful reconstruction of the fashion MNIST dataset, which possesses a higher degree of complexity compared to the handwritten digits, not only validates the DNN's applicability for more challenging scenarios but also suggests its potential in other advanced imaging tasks. Additionally, the consistency in performance across both datasets hints at the network's reliability and its capacity to handle a diverse range of phase images.

Overall, the proposed learning-based phase reconstruction method significantly reduced the computation time for MCF endoscopic lensless imaging, enabling accurate phase reconstruction with an ultra-high frame rate of 181 fps. Furthermore, the proposed DNN can decode the phase of the sample directly from the speckle images and no longer requires the preliminary measurement of the intrinsic phase distortion, which simplifies the imaging process of the fiber endoscopes. Also, the DNN demonstrates accurate phase reconstruction performance with high fidelity. Moreover, the integration of machine learning methodologies could further push the frontiers of endoscopic imaging, opening avenues for more sophisticated diagnostic and therapeutic applications.

\begin{backmatter}
\bmsection{Funding} This work is supported by  Shanghai Artificial Intelligence Laboratory, German Research Foundation (CZ55/40-1), National Key R\&D Program of China (2022ZD0160100) and the National Natural Science Foundation of China (62106183 and 62376222).

\bmsection{Acknowledgments} The authors would like to thank the valuable support from Jiachen Wu, Elias Scharf, Robert Kuschmierz and all the colleagues in MST.

\bmsection{Disclosures} The authors declare no conflicts of interest.

\bmsection{Data availability} The experimentally generated training dataset is publicly available via https://doi.org/10.6084/m9.figshare.24303583.v3

\end{backmatter}

\bibliography{references}

\bibliographyfullrefs{references}


\end{document}